\documentclass[final,3p,times,twocolumn,sort&compress]{elsarticle}
\usepackage{amsmath,bm}
\usepackage{graphicx}
\usepackage{multirow}
\usepackage{epstopdf}
\usepackage{color}
\usepackage{ulem}
\journal{Physics Letters B}

\begin{document}

\title{Parameterization of Deformed Nuclei for Glauber Modeling in Relativistic Heavy Ion Collisions}

\author[label1,label2]{Q. Y. Shou}
\author[label1]{Y. G. Ma} 
\author[label3]{P. Sorensen} 
\author[label3]{A. H. Tang} 
\author[label3]{F. Videb{\ae}k}
\author[label3]{H. Wang} 
\address[label1]{Shanghai Institute of Applied Physics, Chinese Academy of Sciences, Shanghai 201800, China}
\address[label2]{Key Laboratory of Quark and Lepton Physics (MOE) and Institute of Particle Physics, Central China Normal University, Wuhan 430079, China}
\address[label3]{Brookhaven National Laboratory, Upton, New York 11973, USA}

\begin{abstract}

  The density distributions of large nuclei are typically modeled with
  a Woods-Saxon distribution characterized by a radius $R_{0}$ and
  skin depth $a$. Deformation parameters $\beta$ are then introduced
  to describe non-spherical nuclei using an expansion in spherical
  harmonics $R_{0}(1+\beta_2Y^0_2+\beta_4Y^0_4)$. But when a nucleus
  is non-spherical, the $R_{0}$ and $a$ inferred from electron
  scattering experiments that integrate over all nuclear orientations
  cannot be used directly as the parameters in the Woods-Saxon
  distribution. In addition, the $\beta_2$ values typically derived
  from the reduced electric quadrupole transition probability
  B(E2)$\uparrow$ are not directly related to the $\beta_2$ values
  used in the spherical harmonic expansion. B(E2)$\uparrow$ is more
  accurately related to the intrinsic quadrupole moment $Q_{0}$ than
  to $\beta_2$. One can however calculate $Q_0$ for a given $\beta_2$
  and then derive B(E2)$\uparrow$ from $Q_0$. In this paper we
  calculate and tabulate the $R_0$, $a$, and $\beta_2$ values that
  when used in a Woods-Saxon distribution, will give results
  consistent with electron scattering data. We then present
  calculations of the second\ and\ third\ harmonic\ participant\ eccentricity ($\varepsilon_2$ and $\varepsilon_3$)
  with the new and old parameters. We demonstrate that $\varepsilon_3$
  is particularly sensitive to $a$ and argue that using the incorrect
  value of $a$ has important implications for the extraction of viscosity to entropy ratio ($\eta/s$) from the QGP created in Heavy Ion collisions.

\end{abstract}

\maketitle

\section{Introduction}

\label{sec:intro}

In relativistic nucleus-nucleus collisions, the geometry of the
initial overlap region is reflected in the final momentum space
distributions of produced
particles~\cite{Ollitrault:1992bk,Ackermann:2000tr,Adams:2005dq}. How
much that geometry is translated into the final state distributions is
used to infer information about the properties of the matter created
in the collision fireball like its viscosity~\cite{Teaney:2003kp}. The
initial geometry plays a particularly important role in interpreting
the data and in extracting the viscosity to entropy ratio $\eta/s$. It is important therefore to
understand the initial conditions including the exact shape of the
colliding nuclei. The inference of the properties of the fireball from
data is hindered by uncertainties in the characteristics of the
initial state~\cite{Hirano:2005xf}. Recently collisions between
Uranium nuclei (${}^{238}$U) which have an intrinsic prolate
shape~\cite{Raman:1201zz}, have been used as a way to manipulate this
initial geometry in order better test our understanding of the initial
state of heavy ion collisions and the subsequent
fireball~\cite{UUpapers,hardprobes}.

An important part of describing the initial conditions is to correctly
model the geometry of the incoming nuclei. For many years in
simulations for heavy ion collisions, nuclei were approximated as
smooth density distributions and the only anisotropies considered in
the initial state were the intrinsic almond shape caused by the
overlap of two spherical nuclei. As the accumulation of RHIC data
gradually demonstrated that final state anisotropies were sensitive to
the initial geometry and its fluctuations, it became necessary to take
into account the lumpiness of the colliding nuclei
\cite{Miller:2003kd,Alver:2008zza,Bhalerao:2006tp,Broniowski:2007ft,Sorensen:2011hm,Agakishiev:2011eq}.
This is done through Monte-Carlo simulations (M-C) where each nucleus is
generated with a finite number of nucleons distributed with a density
$\rho$ described by a Woods-Saxon distribution~\cite{Miller:2007ri}:
\begin{equation}
\rho(r) = \frac{\rho_0}{1+e^{(r-R_0)/a}},
\label{eq:stdWS}
\end{equation}
where $\rho_{0}$ is the density at the center of the nucleus. The
nuclear radius $R_{0}$ and skin depth $a$ are commonly taken from
high-energy electron scattering measurements~\cite{exp_par_1}. For
non-spherical nuclei, this description was extended by introducing
spherical harmonics in the Woods-Saxon distribution to describe the
modulation of the nuclear radius with $\theta$~\cite{Filip:2009zz}:
\begin{equation}
\rho(r, \theta) = \frac{\rho_0}{1+e^{(r-R_0-R_0\beta_{2} Y_{20}(\theta)-R_0\beta_4Y_{40}(\theta))/a}},
\label{eq:defWS}
\end{equation}
where $\beta_2$ and $\beta_4$ are the deformation
parameters. $\beta_2$ is often derived from measurements of the
reduced electric quadrupole transition probability B(E2)$\uparrow$
from the 0$^{+}$ ground state to the first 2$^{+}$
state~\cite{Raman:1201zz} according to the formula:
\begin{equation}
\beta_{2} = \frac{4\pi}{3ZR^{2}_{0}} \sqrt{\frac{B(E2)\uparrow}{e^{2}}}.
\label{eq:beta2}
\end{equation}
$R_{0}$ is taken to be 1.2A$^{1/3}$. The B(E2)$\uparrow$ values are
directly measured experimental quantities, but the derivation of
$\beta_2$ from B(E2)$\uparrow$ is usually done with model dependent
assumptions including the assumption that the nuclear charge
distribution is a hard edged, step-function rather than a Woods-Saxon
distribution. In order to check for consistency with the measured
B(E2)$\uparrow$ values, one can vary the $\beta_2$ used in
Eq. (\ref{eq:defWS}) and then calculate the intrinsic quadrupole
moment of the resulting nucleus:
\begin{equation}
Q_{20} = \sqrt{\frac{16\pi}{5}}\int d^{3}r\langle\Psi|r^{2}Y_{20}\rho(\vec{r})|\Psi\rangle
\label{eq:q20}
\end{equation}
and check to ensure that $Q_{20}$ is consistent with the measured
B(E2)$\uparrow$ according to the approximation~\cite{eq5}:
\begin{equation}
B(E2)\uparrow = \frac{5}{16\pi} |eQ_{20}|^{2}.
\label{eq:q2b}
\end{equation}
This procedure ensures that the deformation of the simulated nucleus
is consistent with the measured B(E2)$\uparrow$ values for the given
model describing the density profile of the nucleus.

Another difficulty in characterizing deformed nuclei comes from a
mismatch between the parameters inferred from electron
scattering experiments and the Woods-Saxon parameters used in the
simulation of the nuclear density profile. Electron scattering
experiments probe the spherical part of the density distribution
(characterized by radius $R_0$ and diffuseness $a$), which is
averaged over all orientations of the nucleus.  If the nucleus is not
spherical, then the $R_0$ and $a$ used in Eq. (\ref{eq:defWS}) will
not necessarily correspond to the average radius and average
diffuseness inferred from the electron scattering
experiments~\cite{exp_par_1}. For this reason, it is important to
calculate the $R_0$, $a$, and $\beta_2$ values that when used in
conjunction with Eq. (\ref{eq:defWS}) yield a nucleus that is
consistent with the experimental measurements of B(E2)$\uparrow$,
$R_0$ and $a$. In this paper we present updated $R_0$ and $a$ for
nuclei commonly used in collisions at RHIC or the LHC, ${}^{238}$U,
${}^{208}$Pb and ${}^{197}$Au, as well as $\beta_2$ value for
${}^{238}$U.  We then show calculations of the second and third harmonic participant eccentricity $\varepsilon_2$ and
$\varepsilon_3$ for the new and old parameters. We find that in
addition to the obvious dependence of $\varepsilon_2$ on $\beta_2$,
the ratio of $\varepsilon_3/\varepsilon_2$ is very sensitive to the
diffuseness parameter $a$: $\varepsilon_3$ increases with increasing
$a$ while $\varepsilon_2$ decreases so that
$\varepsilon_3/\varepsilon_2$ is overestimated if $a$ is
overestimated. Since viscous damping decreases the ratio of triangular flow over elliptic flow ($v_3/v_2$), an
overestimate of $\varepsilon_3/\varepsilon_2$ will lead to an
over-estimate of the amount of viscous damping needed to match the
experimental data.  We find therefore that the new values of $a$
presented in this work should lead to a decrease in the value of
$\eta/s$ inferred from model-to-data comparisons.

\section{Parameterization of Deformed Nuclei}
\label{sec:model}

When using a Woods-Saxon distribution with parameters $R_0$, $a$, and
$\beta_2$ to characterize the density distribution of a deformed
nucleus, it is obvious that after allowing for all possible
rotations of the nucleus, the averaged density distribution will still
be accurately described by a Woods-Saxon distribution. We find
however, that for the range of $R_0$, $a$, and $\beta_2$ values
describing typical nuclei, the final density distribution after
averaging over all appropriate rotations is indeed well described by a
Woods-Saxon Distribution but with a different radius $R_0'$ and
diffuseness $a'$. To accurately model a nucleus, $R_0'$ and $a'$
should match the parameters measured in electron scattering
experiments. The larger $\beta_2$ is, however, the more $R_0'$
and $a'$ deviate from the $R_0$ and $a$ used in
Eq. (\ref{eq:defWS}).

\begin{figure}[h]
\center
\includegraphics[width=3.3in]{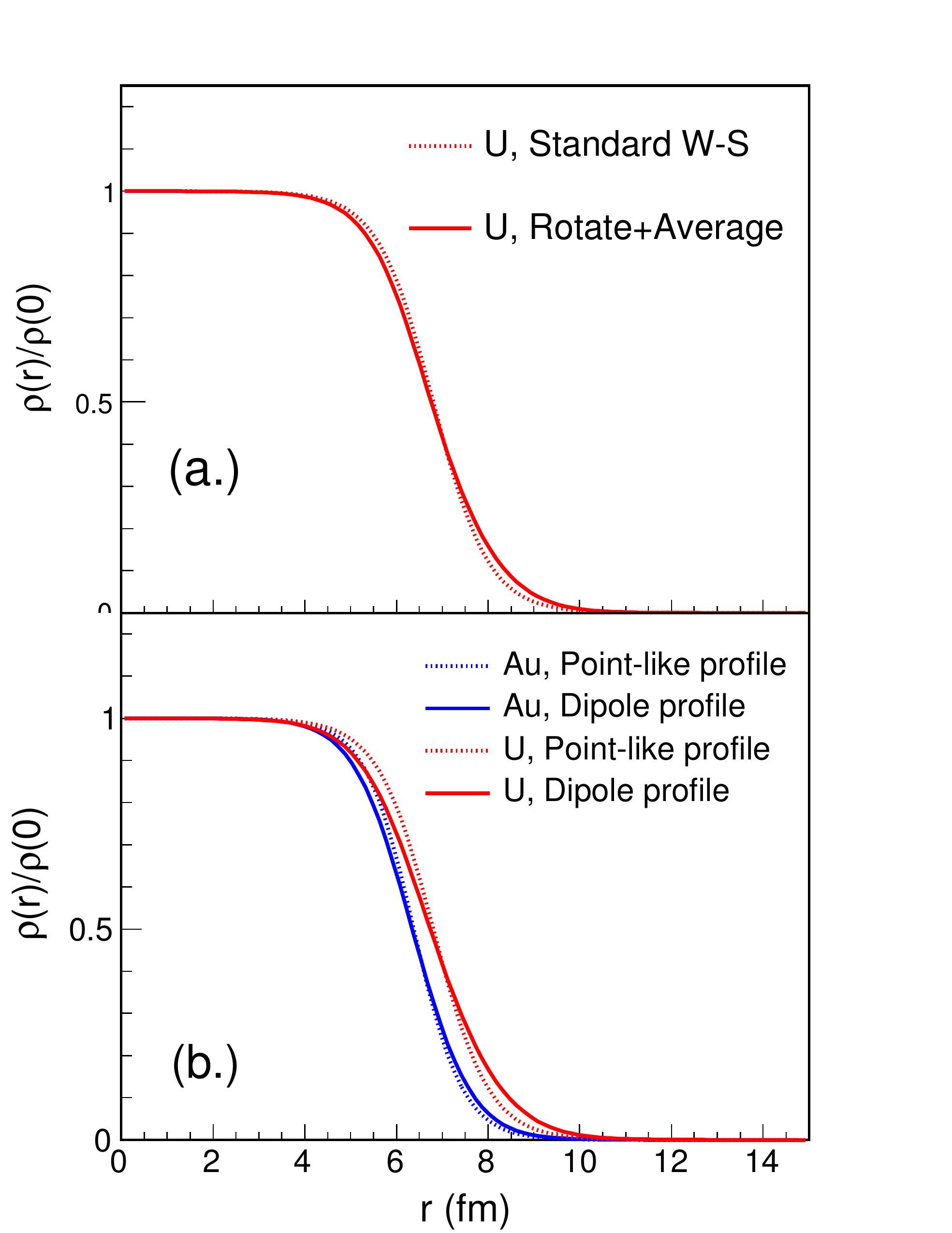}
\caption{(a.) The nucleon density of a deformed nucleus U as a function of
  the radius can be recovered rotating and averaging. (b.) Comparison
  of the nuclear charge density between different nuclei and nucleonic
  profiles.  }
\label{fig:fig1}
\end{figure}

Taking $\rho(r,\theta)$ to be the density distribution of a deformed
nucleus centered at the origin of the spherical coordinate system,
then with n-times randomly rotated and overlapping ellipsoids, the
total nucleon numbers of these n ellipsoids reads:

\begin{equation}
n\iiint \rho(r, \theta) r^2 sin(\theta)\,dr\,d\theta\,d\phi.
\label{eq:ellipInt}
\end{equation}
On the other hand, such total nucleon numbers can also be given by directly
n-times integrating a density of the space, $f$, which only depends on
r:

\begin{equation}
n\iiint f(r) r^2 sin(\theta)\,dr\,d\theta\,d\phi.
\label{eq:sphInt}
\end{equation}
Hence, letting Eq. (\ref{eq:ellipInt}) equal Eq. (\ref{eq:sphInt}), one
can easily identify $f(r)$ as:
\begin{equation}
f(r) = \int \rho(r, \theta) sin(\theta) / 2\,d\theta.
\label{eq:ra}
\end{equation}
Namely, integrating $\rho(r, \theta) sin(\theta) / 2$ over $[0,\pi]$
will give the nucleon density as a function of $r$, as shown in
Fig. (\ref{fig:fig1}) (a.).

In order to correctly reproduce the experimentally measured
B(E2)$\uparrow$ ($Q_{20}$) via Eq. (\ref{eq:q20}) and (\ref{eq:q2b}),
one has to use the charge density distribution to represent the charge
profile of the nucleus as realistically as possible. In this work, we
assume that the charge density of nuclei can be calculated by folding the proton (neutron) density with the charge distribution of the proton (neutron): 
\begin{equation}
\rho_{ch} = \int d^{3}r'[\rho_{p}(r')n_{p}(|r-r'|) + \rho_{n}(r')n_{n}(|r-r'|)],
\label{eq:rhoch}
\end{equation}
where $\rho_{p}$ ($\rho_{n}$) represents proton (neutron) density, while $n_{p}$ ($n_{n}$) is the charge distribution of single proton (neutron). 
It's worth mentioning that the M-C Glauber simulation for the collision of two nuclei does not
typically distinguish neutrons and protons, so we assume both neutrons
and protons follow the same Woods-Saxon distribution in the form of Eq. (\ref{eq:defWS}) for simplicity, however, with different normalization factors:
\begin{equation}
\rho_{p} = \frac{Z}{A}\rho,~~~~~\rho_{n} = \frac{N}{A}\rho,
\end{equation}
where Z, N and A denote proton number, neutron number and nucleon number, respectively.
Based on this assumption, one can utilize B(E2)$\uparrow$ ($Q_{20}$) as a reference to optimize the Woods-Saxon parameters and then apply them to generate
nuclei via a M-C simulation. 


Often, in Eq. (\ref{eq:defWS}), all nucleons are considered as
point-like, viz. $n_{p}$ and $n_{n}$ in Eq. (\ref{eq:rhoch}) are described by delta functions. In a realistic case
however, the finite size of the nucleon should be taken into account
\cite{Hirano2009,Broniowski2009,Rybczyński2014}.  For protons, we adopt the experimentally supported dipole
form of the charge profile \cite{Borkowski1975}, which reads:
\begin{equation}
n_{p}(r) = \frac{m^3}{8\pi}e^{-mr},
\end{equation}
with $m = \sqrt{12}/0.877$ reproducing the proton 
RMS-radius 0.877 fm. For neutrons, we use the following form \cite{Haddad1999}:
\begin{equation}
n_{n}(r) = -\frac{2}{3} \frac{\langle r_{n}^{2} \rangle}{r_{1}^{2} (r_{1} \sqrt{\pi})^3}(\frac{r}{r_{1}})^{2}[1-\frac{2}{5}(\frac{r}{r_{1}})^2]e^{-(r/r1)^2},
\end{equation}
with $\langle r_{n}^{2} \rangle = -0.113$ fm$^2$ and $r_1 = \sqrt{2/5} \cdot 0.71$ fm. B(E2)$\uparrow$ ($Q_{20}$) can then be obtained by using the above euqations. However, our calculation shows that the contribution to $Q_{20}$ from neutrons is very small (much less than 1\%), so we neglect the charge
form factor of neutrons and only take protons into account. In this case, Eq. (\ref{eq:rhoch}) can be simplified to:
\begin{equation}
\rho'(r, \theta) = 2\pi \int\limits^{\infty}_{0} dr' r'^{2} \rho(r',\theta) \int\limits^{1}_{-1} dz n(\sqrt{r'^{2}+r^{2}-2rr'z}),
\label{eq:fws}
\end{equation}
where $r'$ represents the distance from the center of the nucleus
(coordinate origin) to any possible charge points in the proton, while
$r$ represents the distance to the center of the proton. After integrating
Eq. (\ref{eq:fws}) numerically and employing Eq. (\ref{eq:ra}), an
integrated charge distribution of deformed nuclei that only depends on
$r$ can be obtained.

In the steps above, directly adopting parameters from e-A
scattering is no longer valid. Fig. (\ref{fig:fig1}) (b.) shows the charge
density distribution of uranium and gold nuclei with a point-like and
a finite-size profile. The same parameters of $R_{0}$ and $a$ from
Ref. \cite{exp_par_1} are taken for both profiles for each
nucleus. Deviations between the two profiles can been seen from this
figure implying the parameter sets for finite-size profiles need to be
adjusted. A relatively larger radius $R_{0}$ and a smaller diffuseness
$a$ for the finite-size scenario is needed to reproduce the
Woods-Saxon distribution inferred from electron scattering measurements.

Now taking into account the finite size of nucleons and the
presence of a deformation $\beta_2$, we want to find the set of
parameters $\beta_2$, $a$, and $R_0$ that when used in
Eq. (\ref{eq:defWS}) will yield a nucleus consistent with data on
B(E2)$\uparrow$ and electron scattering. During the implementation,
we loop over all possible combinations of $R_{0}$, $a$ and $\beta_{2}$.
For each set, the deformed Woods-Saxon function with finite size profile 
is first rotated and then averaged. After that, the charge distribution
with respect to the radius is extracted. Meanwhile, $Q_{20}$ is directly calculated from
Eq. (\ref{eq:q20}) and (\ref{eq:fws}) in order to check for consistency with
B(E2)$\uparrow$. An optimal set should guarantee that (\romannumeral1) the charge distribution from electron scattering data is reproduced,
(\romannumeral2) the B(E2)$\uparrow$ ($Q_{20}$) value from experiments can be
obtained. For the first criterion, the sum of squares of the residual
($SSR$) is used to quantify the difference between the two
distributions. In this work, only parameter sets providing minimum $SSR$ and the
nearest B(E2)$\uparrow$ ($Q_{20}$) are considered in the final answer. We didn't tune
$\beta_{4}$ because with its typical value,
including $\beta_{4}$ makes little difference to our result.

\section{Results}
\begin{figure}[h]
\center
\includegraphics[width=3.5in]{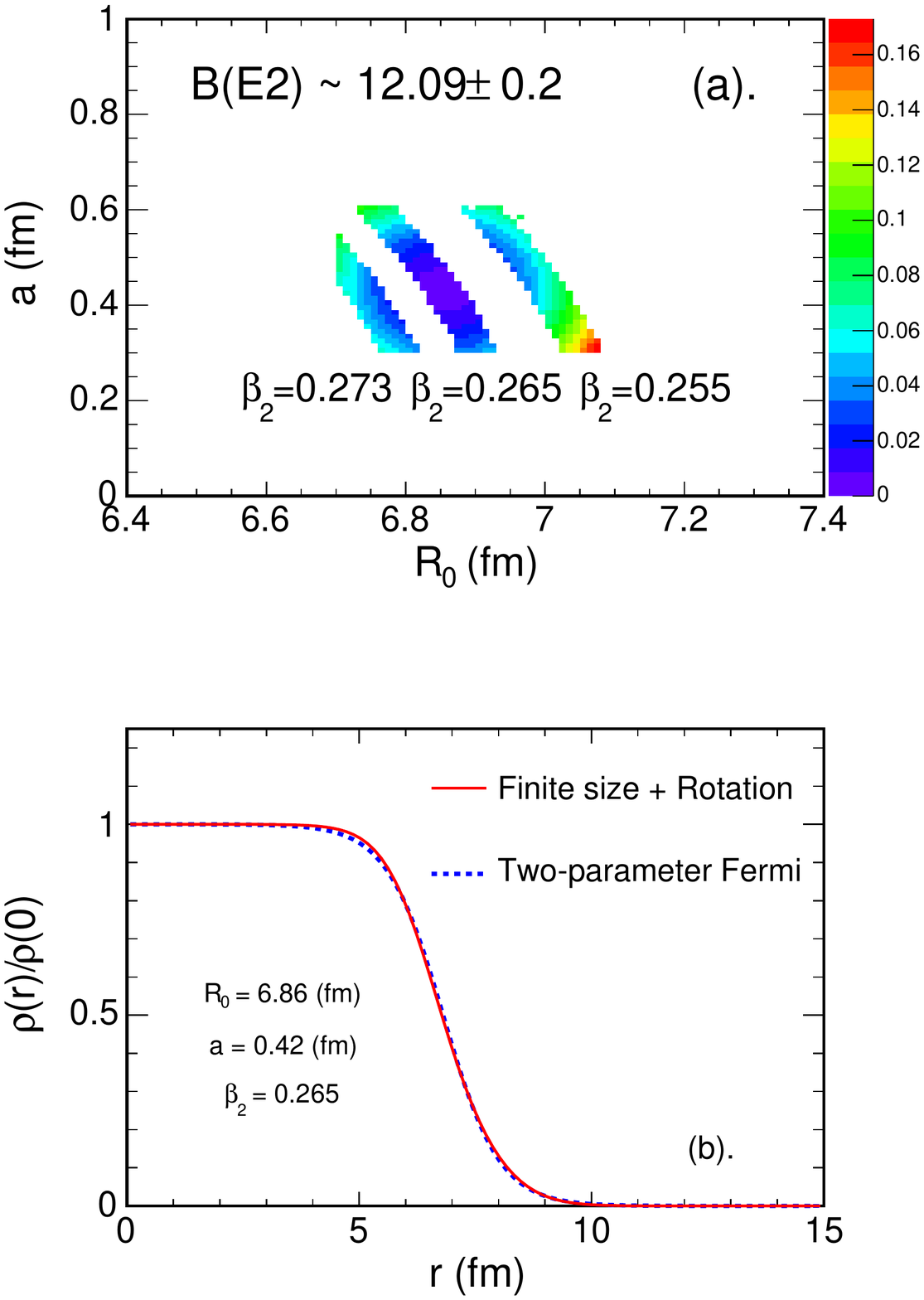}
\caption{(a.) Two dimensional scan of $R_{0}$ and $a$ for ${}^{238}$U
  within the limit of B(E2)$\uparrow$ error. Colors (z axis) represent
  different $SSR$ values. The minimum $SSR$ (purple) can be
  clearly seen at the center of colored region. (b.) The experimentally measured charge distribution (two-parameter Fermi model) has been reproduced by Eq. (\ref{eq:defWS})
  with the optimal parameter set. Here the B(E2)$\uparrow$ value is 12.09.}
\label{fig:scan}
\end{figure}

Fig. (\ref{fig:scan}) shows an example of scanning optimal parameter
sets for ${}^{238}$U. According to Ref. \cite{Raman:1201zz},
the B(E2)$\uparrow$ value of ${}^{238}$U is 12.09$\pm$0.2
($e^{2}b^{2}$). Within this experimental uncertainty, we calculate all
possible combinations of $R_{0}$, $a$ and $\beta_{2}$ to find the
minimum $SSR$. Fig. (\ref{fig:scan}) (a.) shows the two dimension scan
of the parameter sets. The anti-correlation between $a$ and $R_{0}$
is due to the requirement that the calculated B(E2)$\uparrow$ from
this procedure has to match previous measurement of B(E2)$\uparrow$,
of which the uncertainty is reflected as finite band width in the
plot.  The minimum SSR value visible in the plot corresponds to the
purple region centered around the best parameters $R_{0}$=6.86 fm and
$a$ = 0.42 fm. The corresponding $\beta_{2}$ value that gives a nucleus
with a $Q_{20}$ consistent with the B(E2)$\uparrow$ is
0.265. Fig. (\ref{fig:scan}) (b.)  shows that the Woods-Saxon
distribution is recreated by the above parameters with nice consistency
with the one extracted from electron scattering. We summarize the
parameter sets for different nuclear species in
Tab. (\ref{tab:par}). For ${}^{208}$Pb, the ground state nucleus is
considered spherical \cite{Raman:1201zz} although its excited state
may show a deformed configuration, thus the $\beta_2$(${}^{208}$Pb) is
taken to be 0 in this study. Namely, the only contributing step for
${}^{208}$Pb in the aforementioned procedure is the finite-size
correction. For ${}^{197}$Au, B(E2)$\uparrow$ has not been reported from
experiments, therefore $\beta_2$ remains a major uncertainty in the
initial conditions for Au+Au collisions. Therefore we fix the commonly used
value $\beta_2$(${}^{197}$Au)=-0.13, which comes from a mix of data
and model calculations \cite{negBeta2Au}, and only provide the updated
$R_0$ and $a$ values with the caveat that these values will change if
the $\beta_2$ is changed.

\begin{center}
\begin{table}[h]
\caption{Parameter sets for different nuclear species. The parameters
  listed in the "New'' column are the parameters that take into
  account the finite size of nucleons and the deformation of the
  nucleus. "Old" parameters are taken from \cite{Raman:1201zz, exp_par_1}. \newline}  \centering

\begin{tabular}{c | c || c | c | c}
\hline
Nucleus & B(E2)$\uparrow$ & Par. & Old & New \\
\hline
\multirow{3}{*}{${}^{238}$U} &  & $R_{0}$ (fm) & 6.8054 & 6.86 \\
& 12.09 & $a$ (fm) & 0.605 & 0.42 \\
&  & $\beta_2$ & 0.2863 & 0.265 \\
\hline
\multirow{3}{*}{${}^{208}$Pb} &  & $R_{0}$ (fm) & 6.62 & 6.66 \\
& -- & $a$ (fm) & 0.546 & 0.45 \\
&  & $\beta_2$ & 0 & 0 \\
\hline \hline
\multirow{3}{*}{${}^{197}$Au} &  & $R_{0}$ (fm) & 6.38 & 6.42 \\
& -- & $a$ (fm) & 0.535 & 0.41 \\
&  & $\beta_2$ & -- & -0.13 \\
\hline
\end{tabular}
\label{tab:par}
\end{table}
\end{center}

\section{Eccentricity}
To test how the corrections to the Woods-Saxon parameters may affect
observables, we study the multiplicity distribution, the second and third harmonic participant eccentricity ($\varepsilon_2$
and $\varepsilon_3$) in U+U collisions from a Glauber model similar to
the one discussed in Ref.~\cite{Masui:2009qk} with parameters from
Ref. \cite{exp_par_1} and Tab. (\ref{tab:par}).
To generate the multiplicity, we use a two-component model. First
defining
\begin{equation}
n_{AA} = (1-x_{hard})N_{part}/2 + x_{hard}N_{bin},
\end{equation}
where $N_{part}$ is the number of struck nucleons, $N_{bin}$ is the
number of binary nucleon-nucleon collisions, and $x_{hard}$ is a
fractional contribution of $N_{bin}$ to the
multiplicity~\cite{Kharzeev:2000ph,Miller:2007ri}. The multiplicity is
then generated by sampling a negative binomial distribution $n_{AA}$
times with negative binomial parameters ($n_{pp}$ and k) taken from
p+p collisions at the same energy and in the same $|\eta|$
window~\cite{Ansorge:1988kn}. For this calculation we use the
parameters $n_{pp}$= 2.43 and k=2 for the negative binomial and
$x_{hard}=0.13$.
\newline


\begin{figure}[h]
\includegraphics[width=3in]{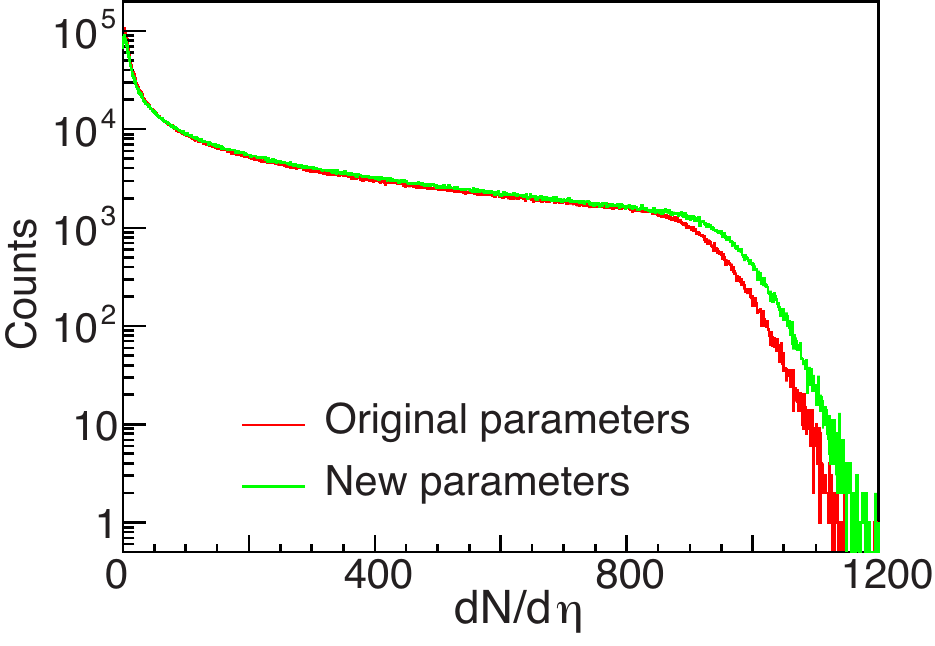}
\caption{\label{fig:dNdeta}(Color online) The $dN/d\eta$ distributions
  in U+U collisions at $\sqrt{s_{NN}}$ = 193 GeV. Two different sets
  of input parameters are compared.}
\end{figure}

Fig. (\ref{fig:dNdeta}) shows the $dN/d\eta$ distributions in U+U
collisions at $\sqrt{s_{NN}}$ = 193 GeV from our Monte Carlo Glauber
model calculations. The maximum $dN/d\eta$ with the new parameters is
slightly higher than that with the original parameters. The increase
comes from the smaller skin depth in the new parameters. As the skin
depth becomes larger, the nucleus becomes more diffuse leading to a
smaller $N_{bin}$ and smaller estimate for $dN/d\eta$.

\begin{figure}[h]
\includegraphics[width=3in]{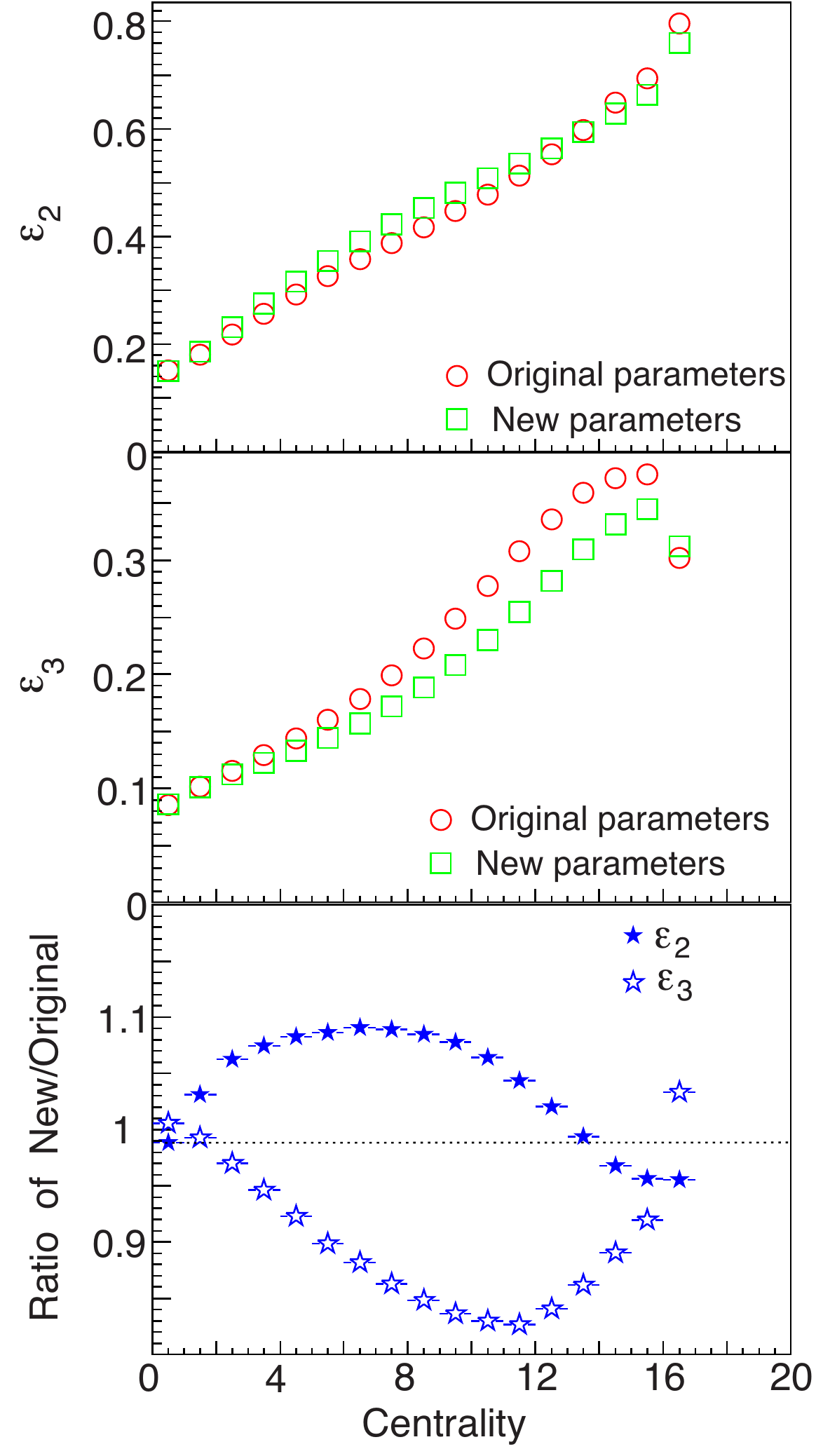}
\caption{\label{fig:eccentricity}(Color online) The second and third
  order participant plane eccentricity as a function of centrality
  bins from Glauber model calculations. Centrality bins 0 to 20
  represent the centrality range 0 to 100\% in 5\% increments with bin
  1 corresponding to 0-5\%.}
\end{figure}

\begin{figure}[h]
\includegraphics[width=3.3in]{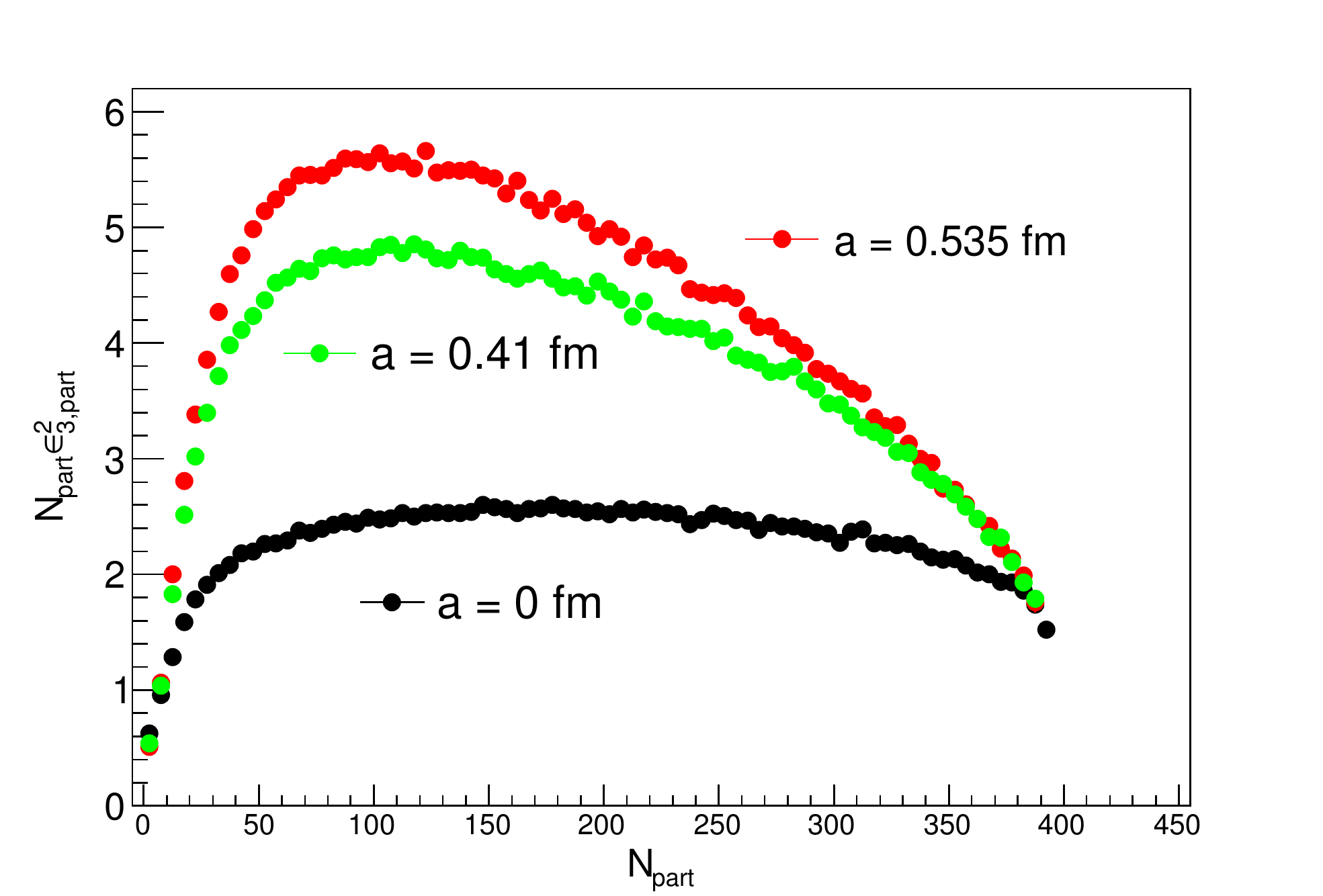}
\caption{\label{fig:a_study} Glauber Model results for
  $N_{part}\varepsilon_3^2$ vs $N_{part}$ for three cases: a diffuse
  nucleus $a=0.41$ fm (new parameters), $a=0.535$ (old parameters), or a hard sphere nucleus $a=0$ fm. }
\end{figure}

To study the impact of the new Woods-Saxon parameters on the initial
geometry, we calculate the second and third harmonic participant
eccentricity. Since different input parameters generate different
multiplicity distributions, we present the eccentricities as a
function of centrality intervals based on the percentage of the total
multiplicity. Fig. (\ref{fig:eccentricity}) upper and middle panels
show $\varepsilon_2$ and $\varepsilon_3$ for the two parameters
sets. The lower panel shows the ratio of the results with the
parameters over the results with old parameters for $\varepsilon_2$
and $\varepsilon_3$. For the most central collisions, both
$\varepsilon_2$ and $\varepsilon_3$ ratios are below one. For the most
central bin, the initial geometry is most sensitive to $\beta_2$ so
the smaller $\beta_2$ values in the new parameter set lead to a
smaller $\varepsilon_2$ and $\varepsilon_3$. In mid-central
collisions, however, the new parameter set produces larger
$\varepsilon_2$ values but smaller $\varepsilon_3$ values. This
behavior can be traced to the smaller value of $a$ in the new parameter
set. In non-central collisions, the value of $\epsilon_3$ is enhanced
by the probability of a nucleon fluctuating out on the edge of one
nucleus and impinging on the center of the other nucleus where it
encounters a relatively large number of nucleons. This effect will be
largest when the nucleon from nucleus A fluctuates in the reaction
plane towards nucleus B. That configuration enhances $\varepsilon_3$
but decreases $\varepsilon_2$. It is this effect that explains the
rise and fall of the ridge correlation in heavy ion
collisions~\cite{Sorensen:2011hm} and it also causes a correlation
between the second and third harmonic event
planes~\cite{Teaney:2010vd}. To further confirm this explanation, in
Fig. (\ref{fig:a_study}) we show a Glauber model calculation of
$N_{part}\varepsilon_3^{2}$ vs $N_{part}$ for Au+Au collisions with
$a$ equal to 0.41 fm (diffuse, new parameter), 0.535 fm (diffuse, old parameter) and 0 fm (hard-sphere). For
mid-central collisions with $N_{part}\approx 100$, the effect of the
diffuseness amplifies $\varepsilon_3$ by a factor of nearly 2.1 (2.5) for new (old) parameters.

From this study, it is clear that the centrality dependence of
$\varepsilon_3$ is strongly dependent on the diffuseness. For that
reason, using the correct value of the diffuseness parameter when
modeling heavy ion collision is crucial especially when calculating
$v_3=\langle\cos(3\phi)\rangle$ where $\phi$ is the azimuth angle of each
particle relative to the major axis of the third harmonic event
anisotropy. Since the relationship of $v_3$ and
$v_2$ similarly defined are often used to estimate the viscosity to entropy ratio
$\eta/s$, estimates of $\eta/s$ from model-to-data comparisons can be
adversely affected if the model does not use the correct Woods-Saxon
parameters.

\section{Summary}
\label{sec:sum}

We find that when modeling the density distribution of deformed nuclei
with a Woods-Saxon distribution, the radius $R_0$ and skin depth
$a$ used in the model will be substantially different than the average
radius $R_0'$ and skin depth $a'$ that would be observed in electron
scattering experiments. The more deformed the nucleus, the larger the
discrepancy. For this reason, one must modify the parameters used in
the Woods-Saxon model so that after appropriately averaging over all
orientations of the axis-of-symmetry for the nucleus, the average
radius $R_0'$ and skin depth $a'$ match the values reported from
electron scattering experiments. In addition, one should also take
into account the radius of the nucleon when carrying out a Monte-Carlo
simulation of the positions of nucleons inside the nucleus, otherwise,
the effective radius will be larger than the simulated radius by
roughly the size of the nucleon. We also find that the model dependent
$\beta_2$ parameters estimated from B(E2)$\uparrow$ measurements, lead
to an overestimate of the deformation of nuclei when they are used in
a deformed Woods-Saxon distribution. We presented a procedure to
calculate the correct values of the Woods-Saxon distribution for
deformed nuclei ($R_0$, $a$, and $\beta_2$) so that the resulting
nucleus is consistent with electron scattering data and
B(E2)$\uparrow$ measurements. Our calculations show that, for
${}^{238}$U, ${}^{208}$Pb and ${}^{197}$Au, the new value of $R_0$ is
slightly larger than the old value while the new value of $a$ is
significantly smaller. For ${}^{238}$U, the obtained $\beta_2$ is also
different than the ones that are commonly used.  We also presented the
results from Glauber-Model Calculations for U+U collisions to study
the effect of the new parameters. The decrease in the skin depth has a
large impact on eccentricity, increasing $\varepsilon_2$ but
decreasing $\varepsilon_3$. Since the amount of viscous damping needed
for a model to match $v_3/v_2$ data will depend strongly on
$\varepsilon_3/\varepsilon_2$, overestimates of $a$ will lead to
over-estimates of $\eta/s$. Predictions of models using the old
parameter values may need to be revisited~\cite{Schenke:2014tga}.

\section*{Acknowledgement}
We would like to thank Thomas Ullrich and Ulrich
Heinz for discussions and suggestions that lead to the initiation of
this study. We thank Thomas Ullrich for the
comments on the paper draft. A. H. Tang thanks G. Wang for discussion
on one of the integrals. This work was partially supported by the
U.S. Department of Energy under contract DE-AC02-98-CH10886, the Major
State Basic Research Development Program in China under Contract
No. 2014CB845401, the National Natural Science Foundation of China
under contract Nos. 11421505, 11035009 and 11220101005, the Key Laboratory of Quark and Lepton Physics (MOE) and Institute of Particle Physics, Central China Normal University under contract Grant No. QLPL2015P02.

{}

\end{document}